\documentclass[12pt]{article}

\usepackage{epsf}
\usepackage[dvips]{graphicx}

\begin{document}

\begin{center}
{\bfseries  Centrality dependence of some characteristics of
relativistic nuclear interactions and percolation cluster
formation}

\vskip 5mm

M.K. Suleymanov$^{1\dag}$, E. U. Khan$^{1}$, K. Ahmed$^{1}$, Mahnaz
Q. Haseeb$^{1}$, Farida Tahir$^{1}$, Ya.H. Huseynaliyev$^{1}$

\vskip 5mm

{\small (1) {\it Department of Physics, COMSATS Institute of
Information Technology, Islamabad}
\\
$\dag$ {\it E-mail: mais@jinr.ru}}
\end{center}

\vskip 5mm

\begin{center}
\begin{minipage}{150mm}
\centerline{\bf Abstract}

Some of the centrality experiments indicate regime change and
saturation in the behaviour of characteristics of the secondary
particles produced in relativistic nuclear interactions. We discuss
 that the responsible mechanism to explain the phenomena could be
the percolation cluster formation and we expect appearance of
deconfinement in the cluster.
\end{minipage}
\end{center}

\vskip 10mm
\section{ Introduction}

The experimental results on the centrality~\cite{[1]}-\cite{[8]} and
energy~\cite{[9]}-\cite{[10]} dependence of particle production in
heavy ion ultrarelativistic collisions  coming from the heavy ion
physics  program at the SPS CERN ~\cite{[11]} and RHIC BNL
~\cite{[12]} show regime change and saturation which could serve as
evidence for the existence of a transition to a new phases of
strongly interacting matter.

Let us consider some  examples. The experimental ratios of  the
average values of multiplicity of  $K^+-, K^--,\phi$- mesons, and
$\Lambda$-hyperons  to  the average values of multiplicity of
$\pi^{\pm}$ - mesons as a function of centrality coming from SPS
CERN are presented in paper~\cite{[1]}. The centrality was fixed
using the average number of participant nucleons ($<N_{part}>$).
They could obtaine that the ratios are increase with centrality and
saturate in the area of $<N_{part}> 60$.

A compilation of measurements of yields at midrapidity for the most
abundant hadron species for central nucleus-nucleus (Au-Au or Pb-Pb)
collisions are presented in paper\cite{[10]}. As the centrality
selection differs between various measurements, they have scaled the
data for the same number of participating nucleons, Npart=350. It is
the point of regime change and saturation effect from these data
too.

The examples show the regime change and saturation in the behaviours
of some characteristics of events  as a function of centrality and
colliding energies.

If the regime change observed in the different experiments  would
take place  unambiguously two times, this would surely be the most
direct experimental evidence seen so far pointing to a phase
transition from the normal hadronic matter to a phase of deconfined
quarks and gluons. However, second point of regime change has not
been observed clearly, moreover the effect taken place for
hadron-nuclear and for nuclear- nuclear interactions too. Let us met
the data coming from the hadron-nuclear experiments.

\section{Hadron-nuclear interaction}

In paper~\cite{[3]} the results are presented from BNL experiment
E910 on pion production and stopping in proton-Be, Cu, and Au
collisions as a function of centrality  at a beam momentum of 18
GeV/c. The centrality of the collisions was characterized using the
measured number of grey tracks, $N_{grey}$, and a derived quantity,
$\nu$, the number of inelastic nucleon-nucleon scatterings suffered
by the projectile during the collision. They obtained that the
$\pi^-$ multiplicity increases approximately proportionally  to
$N_{grey}$  and $\nu$ for all three targets at small values of
$N_{grey}$ or  $\nu$ and saturates  with increasing $N_{grey}$  and
$\nu$ in the region of  more high values of $N_{grey}$  and  $\nu$ .
They could observe that the wounded-nucleon (WN) model~\cite{[3]}
did not explain  the results. It was obtained that the measured
$\Lambda$ yield increased faster than the participant scaling
expectation for  $\nu \le 3$ and then saturated and the deviation in
strange particle production from a wounded-nucleon scaling. The same
result have been obtained by BNL E910 Collaboration for $K^0_s$  and
$K^+$- mesons emitted  in p+Au reaction.

Now let us consider some example on nuclear-nuclear interactions.

\section{Nuclear-Nuclear Interaction}

The average values of multiplicity $<n_s>$ for s - particles
produced in Kr + Em  reactions at 0.95 GeV/nucl  as a function of
centrality are presented in paper~\cite{[5]}. They could saw that
there are two regions in the behaviour of the  values  of  $<n_s>$
as a function of  $N_g$ (a number of grey particles) for the Kr+Em
reaction. In the region of : $N_g  < 40$ the values of $<n_s>$
increase linearly with $N_g$ , here  the  cascade evaporation model
(CEM)~\cite{[13]} also gives the linear dependence but with the
slope  less than the experimental one; $N_g > 40$ the CEM gives the
values for average  $n_s$ greater than  the experimental observed
ones, the last saturates in this region, the effect could not be
described by the model. It have been previously observed in emulsion
experiments ~\cite{[6]}.

It is very important that the regime change has been indicated in
the behaviour of heavy flavor particles production in
ultrarelativistic heavy ion collisions as a function of centrality.
Next paragraph connects with these data.

\section{Regime change and saturation in charmonium production.}

 The ratio of the charmonium  to Drell-Yan cross-sections has been measured by NA38 and NA50 SPS CERN as a function of the centrality of the
 reaction estimated, for each event, from the measured neutral transverse energy $E_t$ ~\cite{[8]}. Whereas peripheral events exhibit the normal
 behavior already measured for lighter projectiles or targets, the charmonium shows a significant anomalous drop of about $20\%$ in the $E_t$  range
 between 40 and 50 GeV.

 Other significant effect which was be seen by authors was a regime change in the $E_t$  range between 40 and 50 GeV both for light and heavy ion
collisions and saturation.

\section{ Results.}

So we could see that:

1. The above motioned regime change had been observed:

- at some values of centrality and colliding energy, as some
critical
     phenomena;

- for hadron-nuclear , nuclear-nuclear interactions and
ultrarelativistic ion collisions;

- in the range energy from  SIS energy  up to RHIC energy;

- almost for all particles (from mesons, baryons,  strange particles
up to charmonium).

2. After point of regime change the saturation is observed.

 3. The simple models (such us WN and CEM) which usually used to describe the high energy hadron-nuclear and nuclear-nuclear interactions could not
explain the existing of the regime change point and saturation.

\section{Discussion}

 The results show that  the dynamic of the phenomena should be same for hadron-nuclear, nuclear-nuclear and heavy ion interactions independent of the
energy and mass of the colliding nuclei and the types of particles.

        The responsible mechanism to describe of the above mentioned phenomena could be statistical and percolation ones because phenomena have a
 critical character. In talk~\cite{[14]} was presented the complicate information about the using statistical and percolation models to explain the
experimental results coming from  heavy ion physics.

       The regime change and saturation  was observed for hadron-nuclear and light nuclear-nuclear interaction where it is very
hard and practically impossible to reach the necessary conditions to
apply the statistical theory (the statistical models have to give
the more strong A-dependences than percolation mechanisms). That is
way we believe that the responsible mechanism for explain the
phenomena could be the percolation cluster formation~\cite{[15]}.

       Big percolation cluster could be formed in the hadron-nuclear, nuclear-nuclear and heavy ion interactions
independent of the colliding energy. But the structure and the
maximum values of the reaching density and temperature of hadronic
matter could be different for different interactions depend on the
colliding energy and masses in the framework of the cluster.

       Paper~\cite{[16]} discusses that deconfinement is expected when the
 density of quarks and gluons becomes so high that it no longer makes sense to partition them into color-neutral hadrons, since these would strongly
 overlap. Instead we have clusters much larger than hadrons, within which color is not confined; deconfinement is thus related to cluster formation.
 This is the central topic of percolation theory, and hence a connection between percolation and deconfinement seems very likely~\cite{[17]}. So we can
 see  that the deconfinement could occur in the percolation cluster. Author explain the charmonium suppression  as a result of deconfinment in cluster
too.

\section{ Appearance of the  critical transparence of
the strongly interacting matter.}

The heavy flavour particles are the most sensitive to phase
transition and to formation of QGP. So the observing of the effects
connected with  formation and decay  of the percolation clusters in
heavy ion collisions at ultrarelativistic energies  and studying the
correlation between these effects and  the effect of  the charmonium
suppression could be unambiguously confirmation of the reaching the
deconfinment  of strongly interacting matter in cluster.

       Percolation cluster is a multibaryon system. With increasing the
centrality of collisions its  size and masses could increase as well
as its absorption capability. So, for example, we could see the
decreasing the particle yields with increasing of the centrality.
But instead it we could see the saturation. It could mean that after
point of regime change the conduction of strongly interacting matter
is increase and it becomes the superconductor. The reason is that
the conduction of the mutter increases and the matter becomes a
superconductor~\cite{[18]} due to the formation of percolation
cluster. Because of in those systems the nucleons (quarks) must be
bound as a result of the percolation.

\section{Search for signal}

       The critical changing of transparency of the strongly interacting matter could influence on the
characteristics of secondary particles changing them.

       Angular  distribution of secondary particles could be more sensitive to the changing of the transparency
of  the matter and  to the formation the big cluster. Because of
the probability of big percolation cluster formation have to be
biggest in the central zone of the collisions. So the particles
which emitted  in different distance from central zone  of
collisions will  be exposed to different changing.   It could lead
to the critical change of the angular correlations of the particle
production because of the transparency the particles will be
absorbed differently depending on the angular. In talk ~\cite{[19]}
$dN\over dy$ distributions for light charged particles ($\pi,p,d,t$)
from Au+Au
 collisions in the range 2 - 8 AGeV at the AGS (E895 Experiment) are presented. They wrote that as collision energy increases, baryons retain
 more and more of the longitudinal momentum of the initial colliding nuclei. This is characterized by a flattening of the invariant particle yields
over a symmetric range of rapidities, about the center of mass, and
it is an indicator of the onset of nuclear transparency.

\section{Summary}

 So, in our opinion  to confirm the deconfinment in cluster it could be necessary to study the centrality dependence of behaviour of  heavy flavour
 particles yield (appearance the point of regime change and  suppression) and simultaneously,  critical increasing of the  transparency of the strongly
interacting matter.

\end{document}